\documentclass[%
 preprint,
 amsmath,amssymb,
 aps,
prl,
]{revtex4-2}

\usepackage{graphicx}
\usepackage{dcolumn}
\usepackage{bm}

\newcommand\njp{New~J.~Phys.}
\let\jpp\undefined
\newcommand\jpp{J.~Plasma~Phys.}
\newcommand\jfm{J.~Fluid~Mech.}
\newcommand\apjl{Astrophys.~J.~Lett.}
\newcommand\apjs{Astrophys.~J.~Suppl.~Series}

\newcommand\ssr{Space~Sci.~Rev.}
\newcommand\mnras{MNRAS}

\newcommand\prx{Phys.~Rev.~X}

\newcommand\aap{A\&A}

\newcommand\frass{Front. Astron. Space Sci.}
\newcommand\frp{Front. Phys.}

\newcommand\bb[1]{\mbox{\boldmath{$#1$}}}
\newcommand\grad{\bb{\nabla}}

\usepackage{scalerel,stackengine}
\stackMath
\newcommand\reallywidehat[1]{%
\savestack{\tmpbox}{\stretchto{%
  \scaleto{%
    \scalerel*[\widthof{\ensuremath{#1}}]{\kern-.6pt\bigwedge\kern-.6pt}%
    {\rule[-\textheight/2]{1ex}{\textheight}}
  }{\textheight}%
}{0.5ex}}%
\stackon[1pt]{#1}{\tmpbox}%
}

\newcommand\reallywidetildeB[1]{
\stackon[-8pt]{#1}{\vstretch{1.5}{\hstretch{1.8}{\widetilde{\phantom{\;\;\;\;\;\;\;}}}}}
}

\newcommand\Bv{\bb{B}}

\newcommand\Jv{\bb{J}}
\newcommand\jv{\bb{j}}

\newcommand\xv{\bb{x}}
\newcommand\vv{\bb{v}}

\newcommand\mi{m_{\rm i}}

\newcommand\uav{\bb{u}_\alpha}

\newcommand\uiv{\bb{u}_{\rm i}}

\newcommand\bPii{\bb{\Pi}_{\rm i}}

\newcommand\whEv{\widehat{\bb{E}}}
\newcommand\whBv{\widehat{\bb{B}}}

\newcommand\wtvarrho{\widetilde{\varrho}}
\newcommand\whuav{\widehat{\bb{u}}_\alpha}
\newcommand\whuiv{\widehat{\bb{u}}_{\rm i}}
\newcommand\whuev{\widehat{\bb{u}}_{\rm e}}

\newcommand\wtbPii{\widetilde{\bb{\Pi}}_{\rm i}}

\newcommand\whbPii{\widehat{\bb{\Pi}}_{\rm i}}

\newcommand\Tnua{\bb{\cal T}_{nu}^{(\alpha)}}
\newcommand\Tnu{\bb{\cal T}_{nu}^{({\rm i})}}
\newcommand\Tnue{\bb{\cal T}_{nu}^{({\rm e})}}

\newcommand\TuxB{\bb{\cal T}_{u\times B}^{({\rm i})}}

\newcommand\Tuu{\bb{\cal T}_{uu}^{({\rm i})}}

\newcommand\whEu{\widehat{\cal E}_{u_{\rm i}}}

\newcommand\whEB{\widehat{\cal E}_B}

\newcommand\whEPii{\widehat{\cal E}_{\Pi_{\rm i}}}

\newcommand\wtPe{\widetilde{P}_{\rm e}}

\newcommand\TPiinablau{{\cal T}_{\Pi\nabla u}^{({\rm i})}}

\newcommand\TJxB{\bb{\cal T}_{J\times B}}

\usepackage{xcolor}

\usepackage[normalem]{ulem}
\usepackage{soul}

\begin{document}

\preprint{APS/123-QED}

\title{Evidence of dual energy transfer driven by magnetic reconnection at sub-ion scales}



\author{Raffaello Foldes}%
 \email{raffaello.foldes@ec-lyon.fr}%
 \affiliation{CNRS, \'Ecole Centrale de Lyon,  INSA de Lyon, Universit\'e Claude Bernard Lyon 1, Laboratoire de M\'ecanique des Fluides et d’Acoustique, F-69134 \'Ecully, France}%
\affiliation{Dipartimento di Scienze Fisiche e Chimiche, Universit\'a dell’Aquila, 67100 Coppito (AQ), Italy}%

\author{Silvio Sergio Cerri}%
 \email{silvio.cerri@oca.eu}%
 \affiliation{Universit\'e C\^{o}te d'Azur, Observatoire de la C\^{o}te d'Azur, CNRS, Laboratoire Lagrange, Bd de l'Observatoire, CS 34229, 06304 Nice cedex 4, France}%

\author{Raffaele Marino}%
 \email{raffaele.marino@ec-lyon.fr}%
 \affiliation{CNRS, \'Ecole Centrale de Lyon,  INSA de Lyon, Universit\'e Claude Bernard Lyon 1, Laboratoire de M\'ecanique des Fluides et d’Acoustique, F-69134 \'Ecully, France}%
 
\author{Enrico Camporeale}%
 \email{enrico.camporeale@noaa.gov}%
 \affiliation{CIRES, University of Colorado \& NOAA Space Weather Prediction Center, Boulder, CO, USA}%

\date{\today}

\begin{abstract}
The properties of energy transfer in the kinetic range of plasma turbulence have fundamental implications on the turbulent heating of space and astrophysical plasmas.
It was recently suggested that magnetic reconnection may be responsible for driving the sub-ion scale cascade, and that this process would be characterized by a direct energy transfer towards even smaller scales (until dissipation), and a simultaneous inverse transfer of energy towards larger scales, until the ion break. 
Here we employ the space-filter technique on high-resolution 2D3V hybrid-Vlasov simulations of continuously driven turbulence providing for the first time quantitative evidence that magnetic reconnection is indeed able to trigger a dual energy transfer originating at sub-ion scales.
\end{abstract}

\maketitle



{\em Introduction.---}Investigations of kinetic-scale plasma turbulence have seen a surge of interest in the past decade, driven by increasingly accurate {\em in-situ} measurements in such range~\cite{AlexandrovaPRL2009,SahraouiPRL2009,ChenPRL2010,BrunoAPJL2015,ChenBoldyrevAPJ2017,WangAPJ2020}. In this context, a transition between magnetohydrodynamic and kinetic regimes occurs when the forward-cascading turbulent energy reaches ion scales \cite{marino2023}. Extensive numerical campaigns have recently been performed in order to better understand the properties of turbulence and plasma heating across and below the so-called {\em ion break}, targeting the interplanetary medium~\cite{FranciAPJ2015,ParasharAPJ2015,CerriAPJL2016,FranciAPJ2016,CerriAPJL2017,GroseljAPJ2017,CerriAPJL2018,GroseljPRL2018,PerronePOP2018,ArzamasskiyAPJ2019,CerriFSPAS2019,GroseljPRX2019,CerriAPJ2021,PassotJPP2022,SquireNatAs2022}. Based on these simulation results, it has been speculated that {\em magnetic reconnection} might be at the origin of the observed ion-break formation driving the subsequent sub-ion scale cascade~\cite{CerriCalifanoNJP2017,FranciAPJL2017}. Such conjecture has been supported both by theoretical arguments~\cite{BoldyrevLoureiroAPJ2017,LoureiroBoldyrevAPJ2017,MalletMNRAS2017,MalletJPP2017} and, at least partially, by solar-wind observations~\cite{VechAPJL2018}. Since then, {\em tearing-mediated turbulence} has been the subject of thorough numerical investigations~\cite{DongPRL2018,PapiniAPJ2019,TeneraniVelli2020,BorgognoAPJ2022,DongSciAd2022,CerriAPJ2022}.
Yet, the role of reconnection in the energy transfer across and below the ion scales remains rather elusive. 
As we show in this letter, an effective approach to tackle potentially relevant transfer mechanisms is provided by the so-called {\em space-filter technique}, originally developed in the context of hydrodynamics for `large-eddy simulations'~\cite{germano1992,MeneveauKatzARFM2000}, and later on adopted as an investigative tool in plasma turbulence~\cite{aluie2010,MorelPOP2011,MieschSSRv2015,YangPRE2017,YangPOP2017,CamporealePRL2018,CerriCamporealePOP2020,AdhikariPRE2021,AlexakisChibbaroJPP2022,ArroAA2021,ManziniPRL2023}.
A qualitative picture of the kinetic-range energy transfer in a tearing-mediated scenario was suggested in~\citet{FranciAPJL2017} (see their Fig.4). 
In that work, the interaction between large-scale vortices feeding the formation of strong current sheets at their boundaries, quickly destroyed by the plasmoid instability, was interpreted as a non-local transfer of energy from the large scales (of the vortices) directly to sub-ion scales (comparable to the thickness of the current sheets). 
Moreover, the continuous formation of small-scale magnetic islands ({\em plasmoids}) and their subsequent merging to form bigger structures was interpreted as an inverse transfer towards larger scales. Thus, a dual transfer of energy should develop at sub-ion scales:
a direct transfer of reconnection-induced fluctuations towards smaller scales until dissipation, and a simultaneous inverse transfer towards the ion break due to the plasmoid growth by island coalescence. This picture has been purely qualitative until now, the presence of bi-directional energy transfers being rigorously assessed in mangnetohydrodinamics plasmas \cite{alexakis2018} as well as in rotating-stratified geophysical fluids \cite{marino2013,marino2015,balwada2022,alexakis2024} .

In this Letter, 2D3V hybrid-kinetic simulations of forced turbulence are analyzed by means of space-filter technique, which allows to investigate the (local and non-local) energy transfer in kinetic plasmas through scales as a function of spatial location and time. We show how the occurrence of magnetic reconnection: (i) enables a consistent energy transfer below ion scales, and (ii) drives a dual (inverse and direct) transfer within the sub-ion range. 


\begin{figure*}[tbh]
\centering
\includegraphics[width=\textwidth]{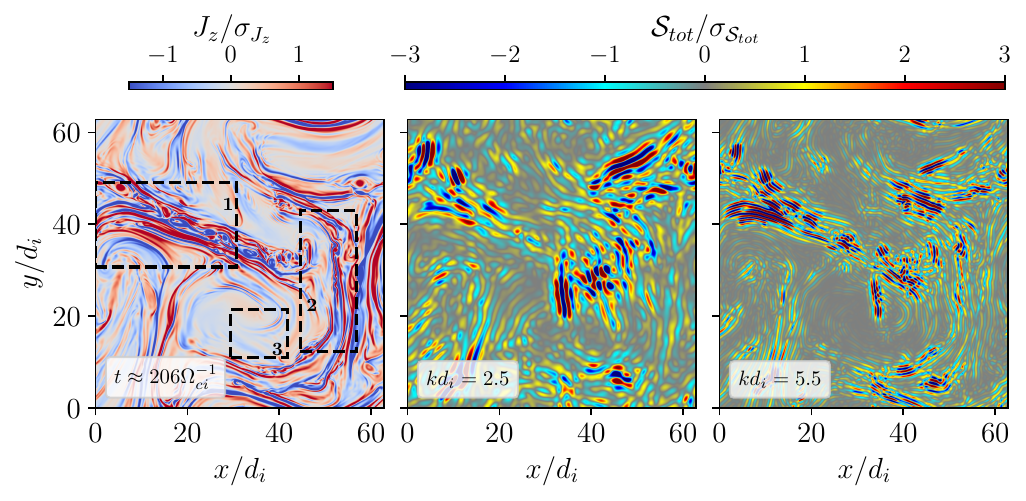}
\caption{({\it Left}) Out-of-plane current density $J_z^\prime=J_z/\sigma_J$ at the time $t\approx206\Omega_{\rm c,i}^{-1}$, when the system reaches a quasi-steady state. ({\it Middle}) Total  energy transfer $\mathcal{S}_{tot}^\prime=\mathcal{S}_{tot}/\sigma_\mathcal{S}$ computed at the scales $kd_i=2.5$ ({\it left}) and $kd_i=5.5$ ({\it right}). The dashed boxes 1 and 2 highlight regions with intense reconnection, box 3 is a reference region in which reconnection is absent or very weak.}
\label{fig:fig1}
\end{figure*}
{\em Method.---}We analyze the 2D3V hybrid-Vlasov-Maxwell (HVM) simulation of continuously driven turbulence in a $\beta_{\rm i}=\beta_{\rm e}=1$ plasma presented in~\citet{CerriCalifanoNJP2017}. The HVM model evolves fully kinetic ions, solving the Vlasov equation for their distribution function $f_{\rm i}(\xv,\vv,t)$, and fluid electrons through a generalized Ohm's law in the quasi-neutral approximation $n_{\rm i}=n_{\rm e}\doteq n$ (displacement current in the Amp\'ere's law is neglected). 
The simulation size is $1024^2$ grid points in real space, spanning a wavenumber range $0.1\leq k d_{\rm i}\leq 51.2$, where $k=k_\perp=(k_x^2+k_y^2)^{1/2}$ and $d_{\rm i}$ is the ion inertial length. An external forcing in the Vlasov equation continuously injects ion-momentum fluctuations at scales $0.1\leq k_{\rm ext} d_{\rm i}\leq0.2$; the magnetic field is initialized (at $t=0$) with small-amplitude perturbations $\delta\bb{B}$ at wavenumbers $0.1\leq k_{\delta B\,}d_{\rm i}\leq0.3$, reaching $\delta B_{\rm rms}/B_0\sim0.1$ in the quasi-steady state.
Results from this HVM simulation were used to conjecture about the existence of a sub-ion-scale tearing-mediated range~\cite{CerriCalifanoNJP2017}; they were later accompanied by a hybrid-PIC simulation to confirm such conjecture~\cite{FranciAPJL2017}. 
Despite fluctuations' properties have been thoroughly analyzed~\cite{CerriJPP2017}, a detailed analysis of the turbulent energy transfer based on this high-resolution numerical simulation had to await the development of proper space-filter formalism and diagnostics for hybrid-kinetic models~\cite{CerriCamporealePOP2020}.\\
In the following, a filtered vector field $\widetilde{\bb{V}}(\bb{x},t)$ denotes the convolution of $\bb{V}(\bb{x},t)$ with a filter $\varphi$, i.e., $\widetilde{\bb{V}}(\bb{x},t)\doteq\int_{\Omega} \varphi(x-\xi)\bb{V}(\bb{x},t){\rm d}\xi$ over the domain $\Omega$. 
Here, we adopt the low-pass Butterworth filter, which in Fourier space reads $\varphi_k=1/[1+(k/k_*)^8]$ with $k_*$ ($\sim\ell_*^{-1}$) being the characteristic filtering wavenumber (scale). The Favre filter of $\bb{V}$ is $\widehat{\bb{V}}\doteq\widetilde{\varrho\bb{V}}/\widetilde{\varrho}$, where $\varrho$ is the mass density.
Filtered equations for the energy channels in general quasi-neutral hybrid-kinetic models are presented in~\cite{CerriCamporealePOP2020}. When dissipation and external injection can be neglected in the HVM model with massless, isothermal electrons, these equations read
\begin{align}
    \frac{\partial\langle\whEu\rangle}{\partial t}\, =\, &
 \,\langle\widehat\Phi_{u_{\rm i},B}\rangle\,
 +
 \,\langle\widehat\Phi_{u_{\rm i},\Pi_{\rm i}}\rangle\,
 -
 \,\langle{\cal S}_{u_{\rm i}}\rangle\,,\label{eq:ionkin-filtered-avg}\\
    \frac{\partial\langle\whEPii\rangle}{\partial t}\, =\, & 
  \,-
  \langle\widehat\Phi_{u_{\rm i},\Pi_{\rm i}}\rangle\,
  -
  \,\langle{\cal S}_{\Pi_{\rm i}}\rangle\,,\label{eq:iontherm-filtered-avg}\\
  \frac{\partial\langle\whEB\rangle}{\partial t}\,=\, & 
  \,\langle\widehat{\cal I}_{\rm e}\rangle\,
  -
  \,\langle\widehat\Phi_{u_{\rm i},B}\rangle\,
  -
  \,\langle{\cal S}_B\rangle\label{eq:magn-filtered-avg}\,,
\end{align}
where $\langle\dots\rangle$ denotes a spatial average, $\whEu=\frac{1}{2}\wtvarrho|\whuiv|^2$, $\whEPii=\frac{1}{2}{\rm tr}[\whbPii]$, and $\whEB=|\whBv|^2/8\pi$ are the ion-kinetic, ion-thermal, and magnetic energy densities at scales $\ell\geq\ell_*$, respectively ($\bPii$ is the ion-pressure tensor and $\uiv$ is the ion-bulk flow, both obtained as $\vv$-space moments of $f_{\rm i}$). 
The injection-like term $\widehat{\cal I}_{\rm e}\doteq \wtPe(\grad\cdot\whuev)$ involving scales $\ell\geq\ell_*$ is due to the isothermal-electron condition $P_{\rm e}=nT_{\rm 0,e}$. 
The terms $\widehat\Phi_{u_{\rm i},B}\doteq \widehat{\jv}_{\rm i}\cdot\whEv$ (where $\widehat{\jv}_{\rm i}=e\widetilde{n}\whuiv$) and $\widehat\Phi_{u_{\rm i},\Pi_{\rm i}}\doteq\wtbPii:\grad\whuiv$ represent energy exchange (i.e., conversion) between different channels (occurring at scales $\ell\geq\ell_*$). Finally, the source/sink terms representing the (local and non-local) energy transfer between large ($k<k_*$) and small ($k>k_*$) scales through the filtering scale $k_*\sim\ell_*^{-1}$ are 
\begin{align}
    {\cal S}_{u_{\rm i}}\, \doteq\, &
 \,\widehat{\jv}_\mathrm{i}\cdot\bb{\epsilon}_\mathrm{MHD}^*\,
 -\,\Tuu:\grad\whuiv\,,\label{eq:S_ui-def}\\
    {\cal S}_{\Pi_{\rm i}}\, \doteq\, & 
  \,\TPiinablau\,,\label{eq:S_PIi-def}\\
  {\cal S}_B\, \doteq\, &
  \,\widehat{\jv}_\mathrm{e}\cdot(\bb{\epsilon}_\mathrm{MHD}^*+\bb{\epsilon}_\mathrm{Hall}^*)\,
  +\,\jv^*\cdot\whEv\label{eq:S_B-def}\,,
\end{align}
where $\widehat{\jv}_{\rm e}=-e\widetilde{n}\whuev=\widetilde{\bb{J}}-e\widetilde{n}\whuiv$ (with $\widetilde{\bb{J}}=\frac{c}{4\pi}\nabla\times\widetilde{\Bv}$), and we have introduced the ``turbulent'' electric fields and current density at scales $\ell<\ell_*$, $\bb{\epsilon}_{\mathrm{MHD}}^* = -\TuxB$, $\bb{\epsilon}_{\mathrm{Hall}}^* = -\TJxB$, and $\jv^* =\Tnu-\Tnue=\widehat{\Jv}-\widetilde{\Jv}$.
The sign convention is such that ${\cal S}>0$ denotes direct energy transfer from large to small scales, while ${\cal S}<0$ means inverse transfer from small to large scales.
The ``sub-grid'' terms ${\cal T}$ associated to nonlinearities are given by $\Tuu\doteq\widetilde{\varrho}(\widehat{\uiv\uiv}-\whuiv\whuiv)$, $\TuxB\doteq\frac{1}{c}(\reallywidehat{\uiv\times\Bv}-\whuiv\times\whBv)$, $\TPiinablau\doteq\reallywidehat{\Pi_{{\rm i},jk}\partial_k u_{{\rm i},j}}-\widetilde{\Pi}_{{\rm i},jk}\partial_k \widehat{u}_{{\rm i},j}$, $\TJxB\doteq\frac{\mi}{ec}\frac{1}{\wtvarrho}(\reallywidetildeB{\Jv\times\Bv}-\widetilde{\Jv}\times\widetilde{\Bv})$, and $\Tnua\doteq\widehat{n\uav}-\widetilde{n}\whuav$.
The corresponding equation for the (filtered) total energy $\widehat{\cal E}=\whEu+\whEPii+\whEB$ is $\partial_t\langle\widehat{\cal E}\rangle=\langle\widehat{\cal I}_{\rm e}\rangle-\langle{\cal S}_{\rm tot}\rangle$, where ${\cal S}_{\rm tot}={\cal S}_{u_{\rm i}}+{\cal S}_{\Pi_{\rm i}}+{\cal S}_B$.
In the following, we focus our analysis on the terms ${\cal S}$, representing the actual transfer through scales.
\begin{figure}[!tbh]
\centering
\includegraphics[width=0.7\textwidth]{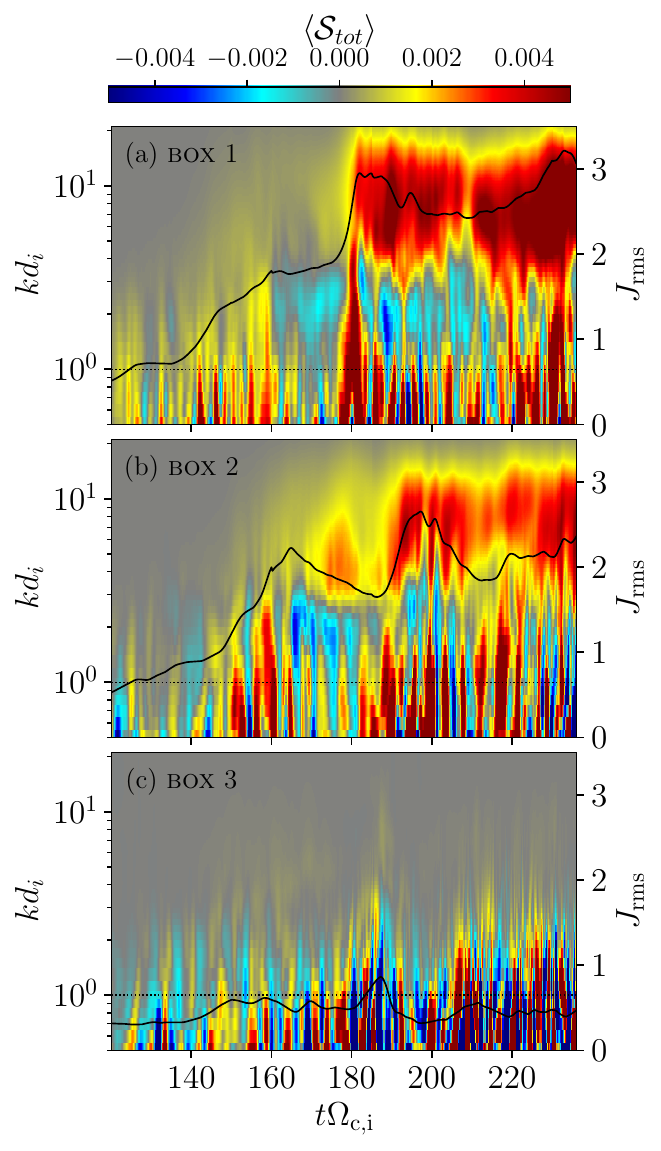}
\caption{Temporal evolution of the total energy transfer averaged $\langle{\cal S}_{\rm tot}\rangle$ over the 3 boxes highlighted in Fig.~\ref{fig:fig1}, in the range $0.5\le kd_i \lesssim 20$. In box 1 and 2 develop the most intense reconnection events, while no intense current sheets can be detected in box 3. Black curves are the root-mean-square current density averaged in the corresponding sub-domains. 
}
\label{fig:fig2}
\end{figure}

{\em Results.---}The simulation exhibits two noteworthy times (in inverse ion-cyclotron frequency units $\Omega_{\rm c,i}^{-1}$): the time of first reconnection events $t_{\rm rec}\approx135$, and the time marking the transition to quasi-steady turbulence $t_{\rm qst}\approx200$ (fig.1 of \cite{FranciAPJL2017}). 
Here we analyze features of the flux terms computed throughout the simulation domain as the plasma dynamics evolve. In order to locate the most prominent reconnection sites, we focus on sub-regions characterized by the highest (on average) values of the density current and the formation of the largest number of plasmoids, indicated as box 1 and box 2 in Fig.~\ref{fig:fig1}. The left panel shows contours of the out-of-plane current density $J_z/\sigma_{J_z}$ at $t\simeq206$, alongside contours of the total-energy transfer ${\cal S}_{\rm tot}/\sigma_{{\cal S}_{\rm tot}}$ (both normalized by their standard deviations) through two representative wavenumbers, $k d_{\rm i}=2.5$ and $k d_{\rm i}=5.5$. To highlight spatial correlations between current structures and total-energy transfer, a movie of Fig.~\ref{fig:fig1} can be found in the supplemental material~\cite{supp_movie}.
These renderings emphasize qualitatively the key result of our analysis: as $k$ increases, the energy transfer becomes significantly less volume filling and more localized within the most intense currents; concurrently, as the time goes by, progressively larger magnetic islands arise from the edge of the current structures~\cite{YangPOP2017,CamporealePRL2018}. These dynamics can be understood as the simultaneous generation of small scales due to the disruption of the (large-scale) current structures by magnetic reconnection, and the nonlinear growth of mesoscale magnetic islands -- corresponding to an upscale energy transfer -- that suggests the existence of a bi-directional energy cascade at sub-ion scales.
In order to quantitavely assess the local-in-space properties of energy transfer as a function of the scale and in time, as well as to characterize the role of magnetic reconnection, we analyze averages of the flux terms computed on box 1 and 2, and on a sub-region pervaded by weaker currents showing no signs of reconnection during the simulation (box 3 in Fig.~\ref{fig:fig1} and in~\cite{supp_movie}). Such analysis is reported in Fig.~\ref{fig:fig2}, where scalograms of the total-energy transfer rate $\langle{\cal S}_{\rm tot}\rangle$, averaged over each box, are plotted as a function of $kd_{\rm i}$ versus simulation time. Red colors in $\langle{\cal S}_{\rm tot}\rangle$ correspond to a (direct) energy transfer from large to  small scales, whereas upscale (inverse) transfer is indicated by blue nuances. The coexistence of forward and inverse transfers -- with a sign inversion occurring in the range $3\lesssim kd_{\rm i}\lesssim4$ -- demonstrates  the existence of a dual cascade, active at sub-ion scales in box 1 and box 2, which is likely triggered by magnetic reconnection. 
The latter is  inferred through the time evolution of the root-mean-square  current density $J_{\rm rms}$ within each box (black lines overlaid on the scalograms). In particular, the dual cascade seems to sets at those times when $J_{\rm rms}$ roughly saturates, $t\simeq160$ and $t\simeq180$ in box 1 and box 2 respectively. Box 3 is instead characterized by a nearly flat  current signal  with relatively low intensity, indicative of  no clear reconnection activity.

\begin{figure*}[!tbh]
\centering
\includegraphics[width=\textwidth]{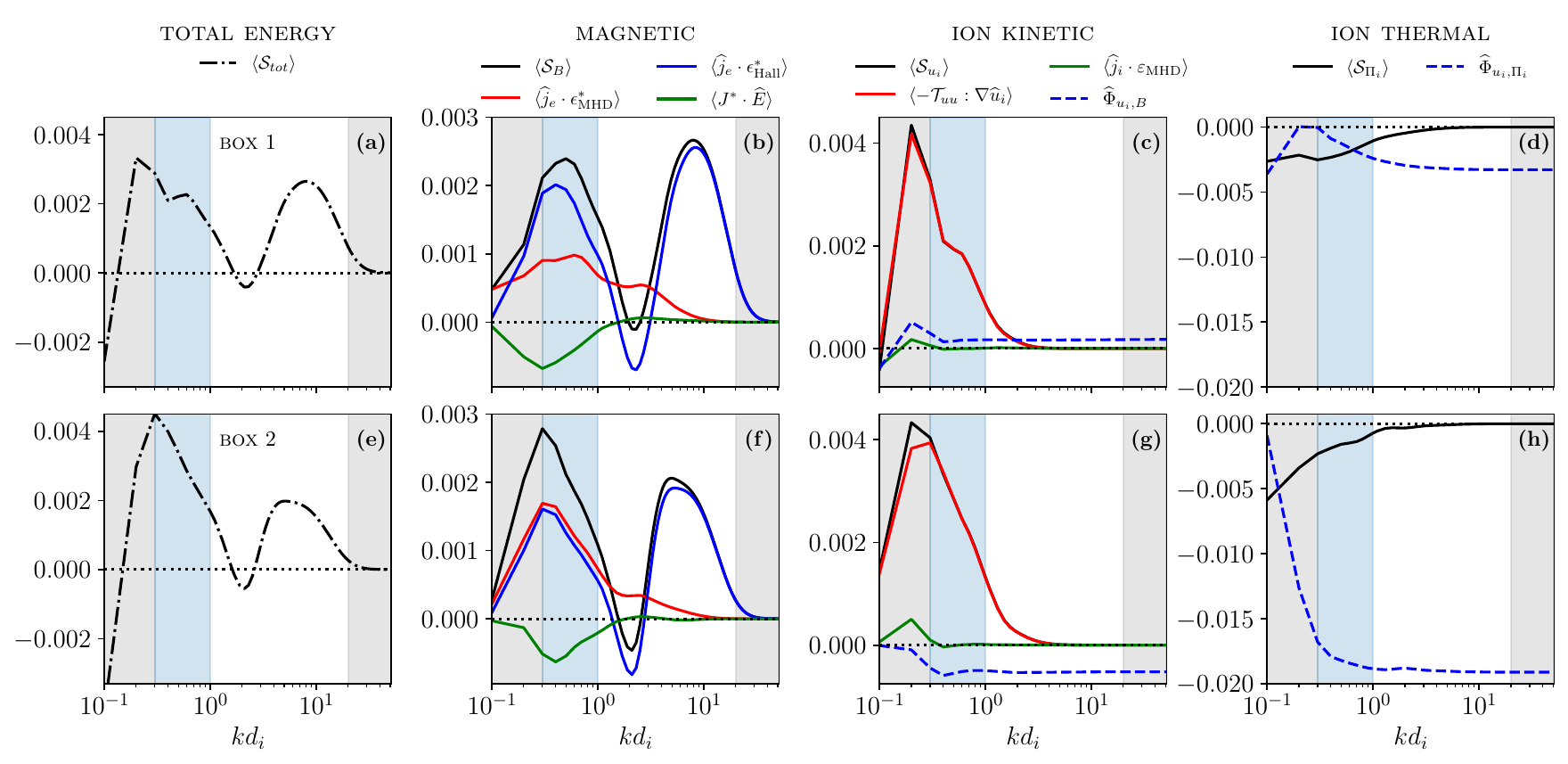}
\caption{Energy transfer terms computed within box 1 and 2 indicated Fig.\ref{fig:fig1} and averaged from $t\approx 165\Omega_{\rm c,i}^{-1}$ to $t\approx 200\Omega_{\rm c,i}^{-1}$. Panels (a)--(d) refers to box 1 and show total, magnetic, ion-kinetic and ion-thermal energy transfer components. The same quantities are displayed in panels (e)--(h) for box 2. Gray-shaded regions denotes the wavenumber ranges affected by the external forcing ($kd_{\rm i}\lesssim0.3$) and by numerical dissipation ($kd_{\rm i}\gtrsim 20$). Blue-shaded areas indicate the MHD range.}
\label{fig:fig3}
\end{figure*}

In this sub-region the sign of $\langle{\cal S}_{\rm tot}\rangle$ switches rapidly between positive and negative values as the time goes by, the most of the total energy transfer being concentrated at scales $kd_{\rm i}\lesssim2$ though with strong oscillations. 
For $t\gtrsim210$, the inverse energy transfer in both box 1 and box 2 becomes more sparse and less intense, which might be due to the presence of constant ion energization/heating processes as system settles to a fully developed turbulent state. As reported by \citet{Lu2023}, a reduction of the reconnection rate may indeed be the consequence of the enhanced ion pressure induced by turbulent forcing. Another possibility is that in our setup the typical timescale over which the system will re-form the current sheets is related to the relatively long eddy turnover time set by the forcing ($\tau_{\rm nl}\sim120\,\Omega_{\rm c,i}^{-1}$); thus intense bursts of reconnection events can occur only on those timescales.
These evidences, including the absence in box 3 of energy transfer at scales $kd_{\rm i}>2$ and no extended segments characterized by definite sign, further support the interpretation that the sub-ion dual cascade is triggered by reconnection events when their intensity attains a certain threshold. 
Scalograms of the channel-specific transfer rates, $\langle{\cal S}_B\rangle$, $\langle{\cal S}_{u_{\rm i}}\rangle$ and $\langle{\cal S}_{\Pi_{\rm i}}\rangle$, averaged over each box, as well as the scalograms of $\langle{\cal S}_{\rm tot}\rangle$ and channel-specific transfer averaged over the entire simulation domain, can be found in the supplemental material~\cite{supp}.

It is worth mentioning that, when averaged over the entire box, in our simulation the dual cascade becomes more intermittent in time, typically emerging only after local peaks in the rms current density; how this feature depends on the system size $L$, on the eddy turnover time set by the forcing and on the outer-scale fluctuations' amplitude is out of the scope of the present work and will require further investigations.

The filtered energy source/sink and exchange terms have been computed point-wise throughout the simulation, then averaged over the boxes indicated in Fig.\ref{fig:fig1} 
within a time interval of roughly $165\Omega_{\rm c,i}^{-1}$ from $t\approx 200\Omega_{\rm c,i}^{-1}$, during which the system is reaching a quasi-steady state. Fig.~\ref{fig:fig3} shows the estimates for box 1, panels (a)--(d), and box 2, panels (e)--(f), in which plasma dynamics are likely driven by magnetic reconnection. The main result of this analysis is reported in the first column, displaying the the total energy transfer $\langle{\cal S}_{\rm tot}\rangle$. 
Two important features emerge from panels (a) and (d): the existence of a simultaneous direct ($\langle{\cal S}_{\rm tot}\rangle>0$) and inverse  ($\langle{\cal S}_{\rm tot}\rangle<0$) transfer occurring at sub-ion scales, in the range $1\lesssim kd_{\rm i}\lesssim20$; a forward energy transfer developing at scales $kd_{\rm i}\lesssim1$, as  expected in the magnetohydrodynamic regime (blue-shaded), downstream of the forcing range (gray-shaded, together with the dissipative range). From panels (b) and (f) one infers that the bi-directional flux of total energy $\langle{\cal S}_{\rm tot}\rangle$ below the ion scale is indeed dominated by the magnetic energy $\langle{\cal S}_B\rangle$. The latter is in turn mostly sustained by  $\widehat{\jv}_\mathrm{e}\cdot\bb{\epsilon}_\mathrm{Hall}^*$, thus by the coupling of the electron currents with the turbulent Hall electric field. The term $\widehat{\jv}_\mathrm{e}\cdot\bb{\epsilon}_\mathrm{MHD}^*$ contributes instead mostly at larger scale, becoming negligible with respect to the Hall term at $kd_{\rm i}\gtrsim5$. 
The term $\jv^*\cdot\whEv$ oscillates around zero in the sub-ion range, becoming non-negligible and negative only for $kd_i\lesssim1$, probably due to the breakup of large-scale current structures by reconnection. On the other hand, negative total and magnetic fluxes at sub-ion scales ($1\lesssim kd_{\rm i}\lesssim4$) are likely associated to the growth of magnetic-islands by coalescence, as anticipated in the previous section (see supplemental material~\cite{supp}).
Panels (c) and (g) reveal how the ion-kinetic energy transfer ($\langle{\cal S}_{u_{\rm i}}\rangle$) dominates only within the MHD regime, where can entirely be accounted for by the coupling of ``sub-grid'' Reynolds stress $\Tuu$ and the large-scale strain tensors $\widehat{\bb{\Sigma}}\doteq\grad\whuiv$,  $\langle{\cal S}_{u_{\rm i}}\rangle\approx\langle\Tuu:\grad\whuiv\rangle$. The latter is marginal at scales $kd_{\rm i}>2$, the transfer of ion-kinetic energy becoming negligible at scales smaller than their gyroradius \citep[as is observed in the ion-flow spectrum, showing a very steep power law at sub-ion scales; see, e.g.,][]{CerriJPP2017,FranciAPJ2018,CalifanoFRP2020}.

Through the whole range of scales resolved, the conversion of magnetic energy $\whEB$ to ion-bulk energy $\whEu$ is driven by the large-scale ion current density interaction with the large-scale electric field, being $\widehat\Phi_{u_{\rm i},B}=\widehat{j}_i\cdot \widehat{E}$, though curves (blue dashed) level off at values with opposite sign, positive for box 1 and negative for box 2.
A remark stemming from the comparison of panels (b-f) and (c-g) is that magnetic ($\langle{\cal S}_B\rangle$) and kinetic ($\langle{\cal S}_{u_{\rm i}}\rangle$) energy transfer have comparable amplitudes in the MHD regime, unlike what happen at scales $kd_i\gg 1$, where $\langle{\cal S}_B\rangle$ dominates.
Finally, panels (d) and (h) show the ion-thermal energy transfer $\langle{\cal S}_{\Pi_{\rm i}}\rangle$ is only a tiny fraction of the total energy flux, peaking around the forcing wavenumber; interestingly conversions between ion-thermal energy and ion-kinetic energy ($\widehat\Phi_{u_{\rm i},\Pi_{\rm i}}$), like $\widehat\Phi_{u_{\rm i},B}$ (panels c and g) saturate at $k d_{\rm i}\gg1$. 
Since the terms $\widehat\Phi(k)$ represent the cumulative conversion up to $k$, their saturation well below the ion scales partly supports the picture that turbulent ion heating mostly occurs at $k_\perp\rho_i\sim1$ and within the first few sub-ion scales~\cite[see, e.g.,][]{SchekochihinAPJS2009,ChandranAPJ2010}.
This scenario is further supported by the trends of derivative of the energy conversion terms (see supplemental material~\cite{supp}).

{\em Conclusions.---}
Exploiting the space-filter techniques, we have shown for the first time, that magnetic reconnection and the consequent island dynamics is associated with (i)  the onset of a quasi-steady turbulent state, and (ii) the emergence of a dual (direct and inverse) transfer of energy originating from sub-ion scales. 
In the case under study, the observed bidirectional energy flux is characterized by 
a sign switch of the total flux $\langle{\cal S}_{{\rm tot}}\rangle$ at around $kd_{\rm i}\sim3$, preceded by another change of sign close to $kd_{\rm i}\sim1$ connecting sub-ion and MHD dynamics. The MHD regime is indeed characterized by a forward energy transfer, driven by the ion-kinetic-energy channel, as expected for a plasma whose the velocity field is forced at large scale. In particular, we found that the dual total energy flux is dominated by the magnetic-energy channel, which driven by the interaction between the large-scale electron-current density and the ``turbulent'' Hall electric field $\widehat{\jv}_\mathrm{e}\cdot\bb{\epsilon}_\mathrm{Hall}^*$).
 
The existence of a simultaneous direct and inverse transfer at sub-ion scales driven by magnetic reconnection may have fundamental implications on our understanding of turbulent ion heating in the solar wind \cite{marino2008,marino2011}, especially in the context of the so-called ``helicity barrier''\cite{PassotSulem2019,MeyrandJPP2021,PassotJPP2022,SquireNatAs2022}.
Moreover, while in our setup the sub-ion-scale dual transfer involves ``ion-coupled'' magnetic reconnection (i.e., reconnection events that develop ion outflows), we believe that an analogous picture would hold also when turbulence is dominated by ``electron-only'' reconnection events~\cite{SharmaPyakurelPOP2019,CalifanoFRP2020,Granier_2024}; we indeed mention that a pre-print addressing a similar issue in the context of electron-only reconnection in merging (sub-ion-scale) flux tubes appeared in \cite{Liu_2024} while we were in the resubmission stage of the present manuscript, further supporting the robustness of our results. In general, we conjecture that a sub-ion-scale dual energy transfer would develop regardless of the micro-physics at play in the reconnecting layer, i.e., independently of the details underlying magnetic reconnection at kinetic scales, provided that the separation between ion scales and the collisionless reconnection scale is large enough.

\begin{acknowledgements}
 R.M. and R.F. acknowledge support from the project ``EVENTFUL" (ANR-20-CE30-0011), funded by the French ``Agence Nationale de la Recherche" - ANR through the program AAPG-2020. E.C. is partially funded by NASA grant 80NSSC20K1275. S.S.C. is supported by the French government, through the UCA$^\text{JEDI}$ Investments in the Future project managed by the National Research Agency (ANR) with the reference number ANR-15-IDEX-01, and by the ANR grant ``MiCRO'' with the reference number ANR-23-CE31-0016.
\end{acknowledgements}

\bibliographystyle{apsrev4-2}
\bibliography{biblio}

\end{document}